\documentclass[12pt,tightenlines,eqsecnum,floats,showpacs,nofootinbib,amsmath,amssymb,aps,prd,
superscriptaddress]{revtex4-1}

\usepackage{graphicx}
\usepackage{amsmath,amssymb}

\usepackage{hyperref}

\def\be{\begin{equation}}
\def\ee{\end{equation}}
\def\ba{\begin{align}}
\def\ea{\end{align}}
\def\f{\frac}
\def\tf{\tfrac}
\def\mC{\mathcal{C}}

\def\H{{\rm H}^\xi}
\def\Ho{\H_{\rm o}}
\def\Hi{\H_{\rm int}}
\def\th{\Theta}
\def\t{\theta}
\def\s{\sigma}
\def\d{\delta}
\def\g{\gamma}
\def\la{\lambda}
\def\ep{\epsilon}

\def\lp{\ell_{\rm Pl}}
\def\rmd{{\rm d}}
\def\sD{\sqrt{\Delta}}

\def\h{\hat}
\def\sgn{{\rm sgn}}

\def\tf{\tfrac}

\begin{document}

\title{Hybrid Quantization: From Bianchi I to the Gowdy
Model}

\author{Mercedes Mart\'in-Benito}
\email{merce.martin@iem.cfmac.csic.es}
\affiliation{Instituto de Estructura de la Materia, CSIC,
Serrano 121, 28006 Madrid, Spain}
\author{Guillermo A. Mena Marug\'an} \email{mena@iem.cfmac.csic.es}
\affiliation{Instituto de Estructura de la Materia, CSIC,
Serrano 121, 28006 Madrid, Spain}
\author{Edward Wilson-Ewing} \email{wilsonewing@gravity.psu.edu}
\affiliation{Center for Fundamental Theory, Institute for
Gravitation and the Cosmos, Pennsylvania State University,
University Park PA 16802, USA}

\begin{abstract}

We complete the quantization
of the vacuum Bianchi I model within the framework of loop quantum cosmology
adopting a new {\sl improved dynamics} scheme put forward recently.
In addition, we revisit the hybrid quantization of the Gowdy $T^3$
cosmologies with linear polarization using that scheme, proving with rigor
some steps that remained unconcluded.
The family of Gowdy $T^3$
cosmologies is an inhomogeneous model whose subset of homogeneous
solutions is given precisely by the vacuum Bianchi I model. Our hybrid approach
combines the new loop quantum cosmology description of this homogeneous sector
with a Fock quantization of the inhomogeneities. Both in the Bianchi I model and
in the Gowdy model the Hamiltonian constraint
provides an evolution equation with respect to the volume of the Bianchi~I
universe, which is a discrete variable with a strictly positive minimum.
We show that, in vacuo, this evolution is
well defined inasmuch as the associated
initial value problem is well posed: physical solutions are completely
determined by the data on the initial section of constant Bianchi I volume.
This fact allows us first to carry out to completion
the quantization of the vacuum Bianchi I model
which had not yet been achieved and then to confirm the
feasibility of the hybrid procedure when the homogeneous sector is
quantized with the new improved dynamics scheme.
\end{abstract}

\pacs{4.60.Pp, 04.60.Kz, 98.80.Qc}

\maketitle

\section{Introduction}
\label{s1}

In order to understand the dynamics of the very early universe,
one must consider quantum gravity phenomena. With this aim,
loop quantum cosmology (LQC) \cite{mb-rev,aa-rev,guille} confronts
the quantization of symmetry reduced models in cosmology following loop quantum
gravity (LQG) methods and ideas \cite{al-rev, cr-book,tt-book}.
Although a full derivation of LQC from LQG has not yet been achieved,
LQC is expected to correctly capture
the behavior of the full theory, at least for those degrees of freedom which are
responsible for the most important features of our universe.

Most of the work to date in LQC has been devoted to the study of the so-called
mini-superspace models, which only contain a finite number of degrees of
freedom. Despite their simplicity, these models are
surprisingly rich inasmuch as they cover many of the situations and
phenomena of interest in LQC and have already
shed light on Planck-scale physics in a cosmological setting.
The study of the flat homogeneous and isotropic cosmological model in LQC
\cite{mb1,abl, aps, acs, kl, mbmmo, vt} has shown that the classical big bang
singularity is resolved by quantum geometry effects and that a quantum bounce
occurs when the matter energy density reaches the critical density
$0.41 \rho_{\rm Pl}$.  All this happens in such a way that the large volume
limit of the model (for semiclassical states) is well approximated by the classical
dynamics dictated by general relativity, just as one should expect.
In addition, further studies have shown that the covariant entropy bound holds
in this model \cite{awe1} and that (assuming the presence of an inflaton
field) the likelihood of obtaining viable initial conditions for slow roll
inflation after the quantum bounce is higher than 99\% \cite{as}.
It has been proven as well that the quantum bounce persists
and that the large volume limit gives the correct classical limit when a
nonzero cosmological constant is present \cite{bp} and also for closed
\cite{apsv, skl} and open \cite{kv} topologies. Finally, although less
is known about anisotropic models as their quantum dynamics are considerably
more complex, it has been shown that the classical singularity is
resolved both in the spatially flat Bianchi type I model
\cite{mbhom,bck,chiou,cv,chi,mbmmp,szulc,awe2} and the spatially curved
Bianchi type II and type IX models \cite{mbhom, bdv, awe3, bdh, we}.

One of the key features in LQC is that the curvature is expressed in terms
of holonomies and one must determine how these holonomies are to be
constructed. For homogeneous and isotropic space-times, it took several years to
understand that one should use the so-called $\bar \mu$-scheme or
{\sl improved dynamics} scheme, presented in Ref.~\cite{aps},
rather than the $\mu_o$-scheme or {\sl old dynamics} scheme,
which had been proposed in Ref.~\cite{abl}. When anisotropic models
were studied, there appeared two natural generalizations to
the $\bar \mu$-scheme, which were presented in Ref.~\cite{cv}.
Since one of these was considerably simpler to work with,
and a priori seemed more natural than the other, a lot of the initial LQC
work studying Bianchi I models followed this procedure until several
problems were pointed out \cite{chiou, szulc, cs}. In particular,
the scaling properties of the (more complicated) alternate procedure
have proven to be more suitable \cite{chi,awe2}. Hence,
unless one allows an explicit dependence of the $\bar \mu$-parameters
on the coordinate cell adopted in the construction of the theory (or
restricts the analysis exclusively to compact topologies where a
distinguished choice of coordinate cell is available), the present
consensus is that this alternate and more complicated procedure is the correct one
to follow. We will refer to these two $\bar \mu$-schemes ---i.e., the original
simpler scheme and the more complicated one recently studied in
Ref.~\cite{awe2}--- as schemes A and B respectively.

The first aim of this paper is to complete the loop quantization of the vacuum
Bianchi I model within scheme B, whose kinematical structure has been
established in Ref.~\cite{awe2} and later analyzed in detail in
Ref.~\cite{mbmm}. We will see that
the Hamiltonian constraint of the model provides a difference equation in an
internal discrete parameter $v$, which is strictly positive and proportional to
the volume of the Bianchi I universe. Employing the form of the
superselection sectors for the anisotropies, which were determined in
Ref.~\cite{mbmm}, we will show here that the quantum evolution equation is indeed
well defined {\em in vacuo}, namely, that one can use $v$ as a ``time''
variable and evolve the wave function in terms of it. In other words, we will
show that a set of initial data, given on the section of minimum $v$,
completely determines the physical solutions. Owing to this fact, we will
be able to obtain the physical structure of the vacuum Bianchi I model
for the first time.

An extra motivation for the consideration of the vacuum Bianchi I model with the
spatial topology of a three-torus is that
its solutions coincide with the subset of homogeneous solutions (homogeneous
sector) of the Gowdy $T^3$ model with linear polarization \cite{gowdy}. Based
on the quantization of vacuum Bianchi I given here, one can then face the quantization of
the Gowdy model in the framework of LQC, allowing for the introduction of
inhomogeneities.

The Gowdy $T^3$ model with linear polarization can be viewed as the simplest
inhomogeneous cosmological model. These cosmological spacetimes admit two
axial Killing vector fields \cite{gowdy} and
they describe universes devoid of matter
which generically start with an initial curvature singularity \cite{mon,ise}.
Their quantization by {\sl standard
methods} has been discussed in detail in the literature
(see, e.g., Refs. \cite{qGow,gowdy-fock}).  It is now well known that
after a complete deparametrization, the model
admits an essentially unique Fock quantization with certain desired properties
\cite{gowdy-fock}.
This explains the attention that has already been paid to the
quantization of the Gowdy $T^3$ spacetimes in the framework of the loop theory
\cite{hybrid1,hybrid2,date1,date2},
with the two-fold purpose of including inhomogeneities in LQC and achieving the
quantum resolution of the singularities of the model. In particular,
Refs.~\cite{hybrid1,hybrid2} succeeded in doing this by proposing a hybrid
quantization which combines the loop quantization of the Bianchi type I
homogeneous sector --where scheme A was adopted for the improved dynamics-- with
the natural Fock quantization of Ref.~\cite{gowdy-fock} for the inhomogeneities.
Since it is generally accepted that scheme A must be replaced with the new scheme
B in the quantization of the Bianchi I model, the second goal of this work is to
discuss and complete the hybrid quantization of the Gowdy model making use
of this alternate scheme for the homogeneous sector.
With this aim, we will put on  a
rigorous basis some steps that are essential for the construction of a well-defined hybrid
quantum model and which were left unfinished in Ref.~\cite{mbmm}.

Even though the kinematical Hilbert space of the hybrid
procedure is naturally separated as the tensor product of two subspaces, namely
the polymer space of the homogeneous sector times the Fock space for the
inhomogeneities, the feasibility of this hybrid procedure is not at all trivial.
As we will see, the Hamiltonian constraint of the model couples both sectors in
a complicated way and it is not obvious whether the constraint can in fact
be promoted to a well-defined operator, especially since the
inhomogeneous sector contains an infinite number of degrees of freedom. Despite
the complexity of the model, it was shown in Refs.~\cite{hybrid1,hybrid2} that
indeed the resulting hybrid quantization can be defined properly for
scheme A and the corresponding physical Hilbert structure was obtained.
Although a priori no relation of this structure with the kinematical one can be presumed,
the standard quantum Fock
description for the inhomogeneities was recovered in fact at the physical level.
Now, the additional difficulties associated with scheme B, where holonomies along
different directions no longer commute, make the new
problem considerably more complicated. In this paper, we will prove that the
scheme B hybrid quantization of the Gowdy $T^3$ model is viable.

Let us mention that, apart from the analyses of the Gowdy model,
other studies in the literature that have investigated the role of
inhomogeneities in the framework of LQC are
Ref. \cite{rv}, which adopts a truncation
of LQG and employs an approximation of the Born-Oppenheimer type, and the
effective analysis of Ref. \cite{bhks}.

The paper is organized as follows. In Sec.~\ref{s2} we recall and summarize the
kinematical structure of the vacuum Bianchi I model with $T^3$ topology
in LQC adopting scheme B. In Sec.~\ref{s3}, first we prove that the notion
of evolution with respect to the volume is well posed as the associated
initial value problem is well defined, and then we complete the
quantization, characterizing the physical Hilbert space and a(n over) complete set of
physical observables. In Sec.~\ref{s4}, using the results obtained for
the Bianchi I model, we show that the quantum dynamics of the Gowdy cosmology
is also well defined when we employ scheme B in its hybrid quantization.
Finally, we conclude in Sec.~\ref{s5} with a discussion of our results and
further comments.

\section{Bianchi I $T^3$ Model in Vacuo: Kinematics}
\label{s2}

This section summarizes the kinematical structure
of the Bianchi I model quantized adopting scheme B for the improved
dynamics prescription. We also include the description of the superselection
sectors. We refer the reader to Refs.~\cite{awe2,mbmm} for more details.

\subsection{Vacuum Bianchi I Hamiltonian Constraint}
\label{s2.1}

In order to describe the classical model,
we choose
angular coordinates $\t, \s, \d \in S^1$ in which the spatial metric is diagonal.
The elementary variables in loop quantum gravity are the Ashtekar-Barbero
connection and the densitized triad. In the Bianchi I model,
owing to homogeneity, each of them can be parametrized
in a diagonal gauge by three coefficients. In terms of the fiducial co-triad
$\{\rmd\t, \rmd\s, \rmd\d\}$ and the corresponding densitized fiducial triad, the
coefficients of the Ashtekar-Barbero connection are given by $c_i/(2\pi)$
and those of
the densitized triad by $p_i/(4\pi^2)$, with $i\in\{\t, \s, \d \}$ and where the
denominators of $2\pi$ come from the periods of our coordinates.
These coefficients form three pairs of canonical
variables, with
\be \label{poisson} \{c_i, p_j\} = 8 \pi G \g \d_{ij}.\ee
Here, $\gamma$ is the Immirzi parameter, $G$ is Newton's constant,
and $\d_{ij}$ the Kronecker delta.

For the study of Bianchi I models in LQC using scheme B, it proves useful to
introduce the variables
\be \la_i = \f{\sgn(p_i)\sqrt{|p_i|}}{(4 \pi \g \sD)^{1/3} \lp} \quad
\mathrm{and}
\quad b_i = \sqrt{\left| \f{p_i \Delta \lp^2}{p_j p_k} \right|} c_i, \ee
where $\lp=\sqrt{G\hbar}$ is the Planck length while the area $\Delta \lp^2$
is the gap in the spectrum of the area operator in LQG.
In these equations, it is understood that repeated indices are not
summed over and that the indices $i, j, k$ are all different.
Note that the above change of
variables is in fact well defined only for nonvanishing triad variables
$p_i$. Later on, we will see that this suffices for the study of the
kinematical arena in LQC.

It is also convenient to introduce the variable
\be v = 2 \la_\t \la_\s \la_\d, \ee
which is proportional to the physical volume of the Bianchi I universe.

The operators that appear in the Hamiltonian constraint of the Bianchi I
model for the considered scheme (see Ref.~\cite{awe2} for details) are $\h\la_i$ and
$\widehat{\sin b_i}$ (or, equivalently, complex exponentials of $b_i$).
It is
easiest to work in the $\la_i$ representation, and in this case $\h\la_i$
acts by multiplication while \cite{note0}
\be \widehat{e^{i b_\t}} \Psi(\la_\t, \la_\s, \la_\d) = \Psi\left(\la_\t
-\f{1}{|\la_\s \la_\d|}, \la_\s, \la_\d\right), \ee
and likewise for the other complex exponentials of the $b_i$'s.
Finally, it is most convenient to change the configuration variables from
$(\la_\t, \la_\s, \la_\d)$ to e.g.~$(v, \la_\s, \la_\d)$, since $v$ behaves in
a simple manner under the action of $\widehat{e^{i b_i}}$, namely
\begin{align} \widehat{e^{i b_\t}} \Psi(v,\la_\s, \la_\d)
&= \Psi\left(v-2\cdot\text{sgn}(\la_\s\la_\d),\la_\s , \la_\d\right),\\
\widehat{e^{i b_\s}} \Psi(v,\la_\s, \la_\d)&=\Psi\left(v-2\cdot\text{sgn}(v\la_\s),
\f{v-2\cdot\text{sgn}(v\la_\s)}{v}\cdot\la_\s , \la_\d\right), \end{align}
where the symbol sgn denotes the sign function.
The action of $\widehat{e^{i b_\d}}$ is obtained by
interchanging the roles of $\la_\s$ and $\la_\d$ in the last equation.
Thus, the effect of
these holonomy operators on the dependence on $v$ is just a constant shift
(up to a sign).

The Hamiltonian constraint operator
obtained with a suitable choice of factor ordering and
after a quantum densitization process \cite{mbmm} (such that
the classical counterpart would
appear in the action multiplied by the densitized lapse
$N_{_{_{\!\!\!\!\!\!\sim}}\;}= N/\sqrt{|p_\t p_\s p_\d|}$)
is given in scheme B by
\be \h{\mC}_H = -\f{1}{16 \pi G \g^2}\Big[\h\th_\t \h\th_\s + \h\th_\s
\h\th_\t +\h\th_\t \h\th_\d + \h\th_\d \h\th_\t + \h\th_\s \h\th_\d +
\h\th_\d \h\th_\s \Big], \ee
where
\be \label{defTh} \h\th_i = \pi \g \lp^2 \widehat{\sqrt{|v|}} \Big[\widehat{\sin (b_i)}
\widehat{\sgn(\la_i)} + \widehat{\sgn(\la_i)} \widehat{\sin (b_i)} \Big]
\widehat{\sqrt{|v|}}. \ee
Note that this is a different factor ordering choice for $\h\th_i$ than
what was chosen in Ref.~\cite{awe2}. This choice is preferable because
the operator has the same action on wave functions supported at large
values of $\la_i$ while its action
for small values of $\la_i$ is much simpler and the consequences are more
transparent. Most importantly, the action of this operator does not allow
any communication between different $(\la_\t, \la_\s, \la_\d)$
octants \cite{mbmmp,hybrid1,mbmm}.

Under the action of $\h{\mC}_H$, the zero volume states decouple.
Therefore, the singularities are resolved kinematically inasmuch as
the quantum states that would correspond to them can be removed
in practice from the kinematical Hilbert space \cite{mbmmp,hybrid1, mbmm}.
Besides, under the action of this Hamiltonian constraint,
the different octants remain invariant. Then, each
octant contains different superselection sectors. This and the fact
that the octants are all related by parity allow us to restrict our
attention to the (strictly) positive octant. We do this for the
remainder of this paper.  Then, the
Hamiltonian constraint acting on a wave function where $v > 4$ gives
\begin{align} \label{qham6} 0 =& \sqrt{v}\Big[(v+2)\sqrt{v+4}\,
\Psi^+_4(v+4,\la_\s,\la_\d) - (v+2)\sqrt v\, \Psi^+_0(v,\la_\s,\la_\d)
\nonumber \\& -(v-2)\sqrt v\, \Psi^-_0(v,\la_\s,\la_\d) + (v-2)
\sqrt{v-4}\,\Psi^-_4(v-4,\la_\s,\la_\s)\Big]. \end{align}
Here $\Psi^\pm_{0,4}$ are defined as follows:
\begin{align} \label{qham7}\Psi^\pm_n(v\pm n,\la_\s,\la_\d)=& \:\Psi
\left(v\pm n,\f{v\pm n}{v\pm2}\cdot\la_\s,\f{v\pm2}{v}\cdot\la_\d
\right)+\Psi\left(v\pm n,\f{v\pm n}{v\pm2}\cdot\la_\s,\la_\d\right)
\nonumber\\& +\Psi\left(v\pm n,\f{v\pm2}{v}\cdot\la_\s,
\f{v\pm n}{v\pm2}\cdot\la_\d\right)+\Psi
\left(v\pm n,\f{v\pm2}{v}\cdot\la_\s, \la_\d\right)\nonumber
\\&+\Psi\left(v\pm n,\la_\s,\f{v\pm2}{v}\cdot\la_\d
\right)+\Psi\left(v\pm n,\la_\s,\f{v\pm n}{v\pm2}\cdot\la_\d\right).
\end{align}
On the other hand, if $2 < v \le 4$, the contribution of $\Psi^-_4$ disappears in Eq.
\eqref{qham6}, whereas if $0 < v \le 2$ the two last contributions in that equation,
namely those proportional to $\Psi^-_4$ and $\Psi^-_0$, are  absent.

\subsection{Superselection Sectors}
\label{s2.2}

It was already pointed out in Ref. \cite{awe2} that there are superselection
sectors in $v$, denoted by a continuous parameter $\ep \in (0, 4]$.
Given such a superselection sector, wave functions only have support on
points where $v = \ep + 4 n$, $n$ being a natural number.

Remarkably, there are also superselection sectors in the $\la_i$'s, although
these sectors have a quite different structure compared to those in $v$.
As it is shown in Ref.~\cite{mbmm}, given $\ep$ and
an initial value $\la_i^\star>0$, the wave function will only have support on
those points that can be expressed in the form
\be \label{sselec} \la_i = \left(\f{\ep - 2}{\ep}\right)^z \prod_k \left(
\f{\ep + 2m_k}{\ep + 2n_k} \right)^{p_k} \la_i^\star, \ee
for some $k$, where $m_k$, $n_k$, and $p_k$ are nonnegative integers,
and $z$ is any integer unless $\ep \le 2$, in which case $z = 0$.
Note that different $\la_i^\star$'s
will yield the same superselection sector if they are related by
Eq.~(\ref{sselec}). It is not difficult to see that one of the available
superselection sectors is the set of nonnegative rational numbers
(it suffices to consider the case $\ep=2$ with $\la_i^\star$
being a rational
number). Furthermore, it follows from Eq.~(\ref{sselec}) that each
superselection sector is countable, dense in $\mathbb{R}^+$,
and that all superselection sectors are isomorphic \cite{mbmm}.

Finally, one can show that the form of the superselection sectors for the
three wave function variables $(v,\la_\s, \la_\d)$ is characterized by
three numbers
$(\ep,\la_\s^\star, \la_\d^\star)$, and is given by the tensor product of the
superselection sectors for each individual variable, that is to say that
there is no restriction on the sector of $\la_\d$ given that of $\la_\s$
or vice versa \cite{mbmm}. In particular, note that if $\la_\s^\star$ and
$\la_\d^\star$ are compatible in the sense of Eq.~(\ref{sselec}),
then the superselection sectors of $\la_\s$ and $\la_\d$
are the same. It is clear that, given one superselection sector
$(\ep, \la_\s^\star, \la_\d^\star)$, a wave function will only have support
on a countable number of points.
Now,
whereas the superselection sector in $v$ only contains
information about a discrete set of points separated by a constant shift,
the superselection sector in the $\la_i$'s encodes the information of
a set of points which are densely distributed in the positive quadrant.

\section{Bianchi I $T^3$ Model in Vacuo: physical structure}
\label{s3}

In this section we will analyze the solutions to the Hamiltonian constraint. We
see from Eq.~\eqref{qham6} that the constraint provides a difference equation
in the parameter $v$, and thus it can be regarded as a (discrete) evolution equation in
this parameter. In the previous analysis of Bianchi I carried out in Ref.~\cite{awe2},
a matter content was added, and the notion of evolution was developed in terms of
a massless scalar field, instead of doing it in terms of this volume
parameter $v$, which has a purely geometric nature. In
the former case, thanks to the suitable properties of the massless field, which is quantized
in a standard Schr\"odinger-like representation, it is
straightforward to prove that the associated initial value problem is
well posed. In fact, in this respect
the situation is quite similar to that found in (relativistic) Quantum Mechanics.
Nonetheless, regarding the geometry part, the physical structure of
the solutions remained unanswered. Now, in the vacuum case considered here, the role of
``time'' is played by the volume variable $v$, which has been polymerically
quantized. Because of its discrete nature, the fact that the associated notion of
evolution is well defined is not trivial. We will show in this section that the dynamics are
correctly posed: a set of initial data evaluated on the section of initial
$v$ completely determines the physical solution. As we will see, the proof is not direct, owing
to the complexity of scheme B. In turn, this result will allow us to
obtain for the first time the physical Hilbert space of the vacuum Bianchi I model in LQC
and a(n over) complete set of
observables, thus completing the quantization of the model.

\subsection{Solutions to the Hamiltonian Constraint}
\label{s3.1}

Since we do not expect generic solutions to the Hamiltonian constraint
to be normalizable in the kinematical Hilbert space,
we will look for solutions in a larger space, namely the algebraic dual of a
suitable
domain of definition for the Hamiltonian constraint operator.
It will be convenient to work
with the variables $x_i = \ln \la_i$ instead of the $\la_i$'s themselves,
as the former variables run over the real line while the latter are
positive and, besides, the $x_i$'s suffer displacements under the action of
the Hamiltonian
constraint operator instead of dilatations or contractions. In fact, from
Eq.~\eqref{sselec}, the superselection sectors
in $x_i$ are formed by those points such that
$(x_i-\ln{\la_i^{\star}}) =w_i \in \mathcal{Z}_\ep$,
where
\be \label{xselec} \mathcal{Z}_\ep = \left\{
z \ln{\left(\f{\ep - 2}{\ep}\right)} +\sum_k \bar{p}_k \ln{\left(
\f{\ep + 2\bar{m}_k}{\ep + 2\bar{n}_k} \right)}\right\}.  \ee
Here, for convenience, we have slightly changed the
notation with respect to Eq.~\eqref{sselec}, so that now
$\bar{m}_k\geq \bar{n}_k$ are nonnegative integers, and $\bar{p}_k$ can
take
any integer value. Recall that
$z$ is any integer unless $\ep\leq 2$ in which case $z=0$.
We note that ${\mathcal{Z}}_\ep$ is dense in the real line, because the
superselection sectors of $\la_i$ are dense in $\mathbb{R}^+$ and
the logarithm is a continuous function from the positive axis to the
real line. In spite of the introduction of the $x_i$'s,
we will still keep $v$ as one of our variables given its
nice behavior under the action of the Hamiltonian constraint operator.

Since the Wheeler-DeWitt equation associated to the Bianchi I model
is actually a first order differential equation in the three variables
$x_i$, it should be possible to determine the entire solution
to the Hamiltonian constraint supplying as initial data its
restriction to one Cauchy slice (i.e., a surface with constant value
of one of the $x_i$'s or alternatively with constant $v$). However,
the constraint in LQC is a second order
difference operator for generic values of $v$. Therefore, it is not
immediately clear how the solution can be determined from one slice
of initial data obtained at a constant value of $v$. The solution
lies in the different form that the action of the Hamiltonian constraint
operator has on states with $v \le 4$, this form is that of
a first order difference equation.

Given a superselection sector in $v$, denoted by $\ep \in (0, 4]$, one
obtains this first order difference equation in $v$ for $v = \ep + 4$ in
terms of the initial data on the slice $v = \ep$.  If one can solve this
(highly coupled) difference equation, it will then be possible to solve
the (again highly coupled) second order difference equation for
$v = \ep + 8$ in terms of the data on the slices $v = \ep$ and
$v = \ep + 4$. One can then follow this strategy in order to obtain
the full solution to the Hamiltonian constraint for all $v$.
Finally, because the $v + 4$ terms always appear in the same combination,
given by $\Psi_4^+$, we only need to show how to derive the data on
$v = \ep + 4$, and then the data for all other $v$ can be obtained in the
same manner.

The difference equation we are interested in is
\be \label{diff} \Psi_4^+(\ep+4,x_\s, x_\d) = \sqrt \f{\ep}{\ep + 4} \Big[
\Psi_0^+(\ep,x_\s, x_\d) + \f{\ep - 2}{\ep + 2} \Psi_0^-(\ep,x_\s, x_\d)
\Big], \ee
where the second term on the righthand side is absent if
$\ep \le 2$. Since the righthand side is known, the question is just whether
one is able to obtain $\Psi(\ep + 4,x_\s, x_\d)$ from
$\Psi_4^+(\ep+4,x_\s, x_\d)$ in order to derive the form of the wave
function for all $v$.

The explicit form of $\Psi_4^+$ is given by:
\begin{align}\label{eq:combinations-v4}
\Psi_4^+(v+4,x_\s,x_\d)
&=\Psi\left(v+4,x_\s,\ln{\left(\f{v+4}{v+ 2}\right)}+ x_\d\right)
+\Psi\left(v+4,\ln{\left(\f{v+4}{v+ 2}\right)}+x_\s, x_\d\right)
\nonumber\\&
+\Psi\left(v+4, x_\s, \ln{\left(\f{v+2}{v}\right)}+x_\d\right)
+\Psi\left(v+4,\ln{\left(\f{v+2}{v}\right)}+ x_\s, x_\d\right)
\nonumber\\&
+\Psi\left(v+4,\ln{\left(\f{v+2}{v}\right)}+ x_\s,
\ln{\left(\f{v+4}{v+ 2}\right)}+ x_\d\right)\nonumber\\&
+\Psi\left(v+4,\ln{\left(\f{v+4}{v+ 2}\right)}+ x_\s,
\ln{\left(\f{v+2}{v}\right)}+ x_\d\right)
\end{align}
for any value of $v$ (and, in particular, for $v=\ep$). It
can be expressed as the result of the action of two separate operators,
\be\label{evol} \Psi_4^+(v+4,x_\s, x_\d) = \h U_6(v+4) \h A \Psi(v,x_\s, x_\d),
\ee
where $\h A $ only shifts the value of $v$, namely,
\be \h A \Psi(v,x_\s, x_\d) = \left( \widehat{e^{-i b_\t}} \right)^2
\Psi(v,x_\s, x_\d) = \Psi(v+4,x_\s, x_\d), \ee
while
\begin{align} \h U_6(v) \Psi(v,x_\s, x_\d) &= \Big[ \widehat{e^{-i b_\t}}
\widehat{e^{-i b_\s}} +
\widehat{e^{-i b_\s}} \widehat{e^{-i b_\t}} + \widehat{e^{-i b_\s}}
\widehat{e^{-i b_\d}} + \widehat{e^{-i b_\d}} \widehat{e^{-i b_\s}}
\nonumber \\ & \quad + \widehat{e^{-i b_\d}} \widehat{e^{-i b_\t}}
+ \widehat{e^{-i b_\t}} \widehat{e^{-i b_\d}} \Big]
\left( \widehat{e^{i b_\t}} \right)^2 \Psi(v,x_\s, x_\d)  \end{align}
has a trivial action on the $v$-sector.

The invertibility of the operator $\h U_6 (v)$ for any value of $v$ would
guarantee that one can
determine $\Psi(v,x_\s, x_\d)$ from $\Psi_4^+(v,x_\s, x_\d)$.
Assuming for the moment that the inverse operator
$[\h U_6(v)]^{-1}$ exists, we can derive the state at the
volume $\ep + 4$ by calculating
\be \Psi(\ep + 4, x_\s, x_\d) = \sqrt \f{\ep}{\ep + 4}
\Big[\h U_6(\ep+4)\Big]^{-1}
\Big[ \Psi_0^+(\ep,x_\s, x_\d) + \f{\ep - 2}{\ep + 2}
\Psi_0^-(\ep,x_\s, x_\d) \Big]. \ee
Once again, the second term in the square brackets on the right hand side
of this equation does not appear if $\ep \le 2$.

Again assuming the existence of $[\h U_6(v)]^{-1}$,
it is now straightforward to obtain the value of the wave function for
the section $v = \ep + 8$:
\begin{align} \Psi(\ep + 8,x_\s, x_\d) =& \sqrt{\f{\ep + 4}{\ep + 8}}\,
\Big[\h U_6(\ep+8)\Big]^{-1}
\Bigg[  \Psi^+_0(\ep+4,x_\s, x_\d)
+ \f{\ep + 2}{\ep + 6} \Psi^-_0(\ep+4,x_\s, x_\d)
\nonumber \\& - \f{\ep + 2}{\ep + 6}\sqrt{\f{\ep}{\ep + 4}}\,
\Psi^-_4(\ep,x_\s, x_\d)\Bigg]. \end{align}
It is clear how to repeat this procedure in order to get the
value of the wave function for all larger $v$ as well. Therefore, if the
operator $\h U_6 (v)$ can be inverted  we conclude
that the initial value problem in terms of $v$ is well posed, at
least at $v=\ep$. In the next section we show that indeed this is the case.

\subsection{The operator $\h U_6$}
\label{s3.2}

Let us then analyze the operator $\h U_6(v)$ to see that its
action can be inverted. First, we provide a suitable domain of
definition for $\h U_6(v)$,
keeping fixed the value of $v$ (in other words, we restrict
the discussion just to a slice of constant $v$).
For each direction $i=\s$ or $\d$, consider the linear span
$\text{Cyl}_{\lambda_i^\star}$ of the states whose support is just
one point $x_i$ of the superselection sector determined by Eq.
\eqref{xselec}, with
$(x_i-\ln{\la_i}^\star)=w_i \in \mathcal{Z}_\ep$.
We call $\mathcal H_{\lambda_i^\star}$
the Hilbert completion of this vector space with the discrete inner product.
Then, we can choose the tensor product $\text{Cyl}_{\lambda_\s^\star}
\otimes \text{Cyl}_{\lambda_\d^\star}$ as the domain for $\h U_6(v)$.

Now, if we define on the Hilbert space
$\mathcal H_{\la_\s^\star}\otimes\mathcal H_{\la_\d^\star}$ the
translations
\be\label{transfg}
\h U^{(w_\s,w_\d)}\Psi(v,x_\s,x_\d)=\Psi(v,w_\s+x_\s,w_\d+x_\d),
\ee
then the operator $\h U_6(v)$
is just a sum of six translations of this kind.
These translation operators are unitary because, if
$w_\s$ and $w_\d$ are two numbers in $\mathcal{Z}_\ep$, so that a shift of
$x_i$ by any of them leaves invariant the superselection sector [see
Eq.~\eqref{xselec}], then the sum of
$|\h U^{(w_\s,w_\d)}\Psi(v,x_\s,x_\d)|^2$ over all $x_\s$ and $x_\d$
in the superselection sector coincides with the sum of
$|\Psi(v,x_\s,x_\d)|^2$. Moreover, owing to this property and
the Schwarz inequality, we conclude that the norm of the operator $\h U_6(v)$ is
bounded by 6.

Since $\h U_6(v)$ is bounded, it can be extended as a well-defined
operator to the entire Hilbert space. This extension [which we also
denote by $\h U_6(v)$] provides in fact a normal operator
---namely, the operator commutes with its adjoint---
as the translations in $x_\s$ and/or in $x_\d$ commute.
Hence, in particular, it is guaranteed that the residual spectrum is
empty. Thus the operator $\h U_6(v)$ is
invertible in our Hilbert space if and only if its point spectrum
does not contain the zero.

It is not difficult to convince oneself that the point spectrum
of $\h U_6(v)$ must be empty owing to the properties of the operator. The idea
is that since this operator is just a linear
combination of translations, any of its eigenfunctions must possess a
certain translational invariance which would prevent them from being normalizable.
In order to see this, let us consider again the translations
$\h U^{(w_\s,w_\d)}$, with $w_\s,w_\d
\in \mathcal{Z}_\ep$.
Since they all commute with each other as well as with $\h U_6(v)$,
they can all be diagonalized simultaneously, that is, there exists
a basis of common (generalized) eigenfunctions. Let us
call in the following
\be \label{transsd}
\h U_\s^{w_\s}= \h U^{(w_\s,0)},\quad \h U_\d^{w_\d}= \h U^{(0,w_\d)},
\ee
so that $\h U^{(w_\s,w_\d)}=\h U_\s^{w_\s} \h U_\d^{w_\d}$. Given one
of the eigenfunctions common to all of these translations, we
denote the corresponding eigenvalue of $\h U_i^{w_i}$ by $\rho_i(w_i)$
(with $i=\s,\d$).  This eigenvalue
must be a complex number of unit norm, because the translation operators
are unitary. In addition, since $\h U_i^{w_i}\h U_i^{\bar{w}_i}=
\h U_i^{w_i +\bar{w}_i}$, it follows that
\be\label{compoeigen}
\rho_i(w_i)\rho_i(\bar{w}_i)=
\rho_i(w_i +\bar{w}_i).
\ee
Recalling that all points in the superselection sector can be reached
from $\ln{\la_i^{\star}}$ by a translation $\h U_i^{w_i}$, it is a
simple exercise to show that the eigenfunctions are proportional to
$\rho_\s(w_\s)\rho_\d(w_\d)$. We can always change this wave function
by a constant of unit norm, and thus we fix $\rho_i(0)=1$.
Besides, in order to determine completely the wave function, we only
need to know the value of $\rho_i(w_i)$ in an appropriate subset of
$\mathcal{Z}_\ep$, namely any collection of noncommensurable points
which can generate the entire set by multiplication by
integers. It is possible to see that property \eqref{compoeigen}
provides then all the information about $\rho_i$ at the rest of points
in $\mathcal{Z}_\ep$. In particular, $\rho_i(n w_i)=[\rho_i(w_i)]^n$.

The wave functions $\rho_i(w_i)$ are clearly nonnormalizable with
respect to the discrete inner product in
$\mathcal H_{\lambda_i^\star}$ as they have complex
unit norm at each point of the superselection sector (the
shift of $\mathcal{Z}_\ep$ by $\ln{\la_i^\star}$)
and the sector contains an infinite number of points. In addition,
different wave functions $\rho_i(w_i)$ must be orthogonal, because there
always exists a (unitary) translation operator on
$\mathcal H_{\lambda_i^\star}$ whose eigenvalue differs
for the two wave functions.

At this stage of the discussion, it is worth noticing that,
by the very construction of the algebra of
fundamental operators in LQC previous to the introduction
of superselection sectors, the operators $\h U_i^{w_i}$
that act as translations in the $x_i$ representation can be
identified in the holonomy/connection representation
---where they act as
multiplicative operators--- as elements of the Bohr
compactification of the real line, $\mathbb{R}_{\text{Bohr}}$
\cite{urdiales}. These elements can be understood as maps $\rho_i$
from the real line (corresponding to all possible real values of
$x_i$, or equivalently of $w_i$) to the circle such that
they satisfy condition \eqref{compoeigen} and $\rho_i(0)=1$.
Owing to superselection, however,
the values of $w_i$ are now restricted to
belong to $\mathcal{Z}_\ep$. We can then identify the wave functions
$\rho_i(w_i)$ as
equivalence classes of elements in $\mathbb{R}_{\text{Bohr}}$,
the equivalence relation being the identification of all those
maps $\rho_i$ which differ only by their
action on the set complementary to $\mathcal{Z}_\ep$ in the real
line, i.e., $\mathbb{R}\setminus\mathcal{Z}_\ep$.
Examples of $\rho_i(w_i)$ are provided by the exponential maps
$\exp{(ik_i w_i)}$ from $\mathcal{Z}_\ep$ to $S^1$.
Since $\mathcal{Z}_\ep$ contains noncommensurable
numbers, these exponentials separate all real values
of $k_i$ [that is, for any two values of $k_i$ one can
find a value of $w_i$ for which the exponentials
$\exp{(ik_i w_i)}$ are different].
So, the set of possible and distinct
$\rho_i$ contains all the exponentials with $k_i\in \mathbb{R}$.

Returning to the operator $\h U_6(v)$, it is straightforward to find
its eigenvalue for each of the analyzed wave functions. It is given by
\begin{eqnarray} \label{eigen2}
\omega_6(\rho_\s,\rho_\d)&=&
\sum_{i=\s, \d}\left\{
\rho_i\left[\ln{\left(\tf{v}{v-2}\right)}\right]
+ \rho_i\left[\ln{\left(\tf{v-2}{v-4}\right)}\right]\right\}
\nonumber\\
&+& \sum_{i,j=\s, \d; i\neq j}\left\{
\rho_i\left[\ln{\left(\tf{v}{v-2}\right)}\right]
\rho_j\left[\ln{\left(\tf{v-2}{v-4}\right)}\right]\right\}.
\end{eqnarray}
Remember that here $v>4$.
The point spectrum of $\h U_6(v)$
will not contain the zero provided that there is
no normalizable linear superposition of the above wave
functions with $\omega_6(\rho_\s,\rho_\d)=0$. In this superposition,
the measure for $\rho_i$ is continuous: this
is a consequence of the wave functions $\rho_i(w_i)$
not being normalizable in $\mathcal H_{\lambda_i^\star}$. The
restriction to the kernel of $\h U_6(v)$ is achieved then by introducing
a delta function of $\omega_6(\rho_\s,\rho_\d)$ (peaked at zero).
If one computes the norm of this superposition, the orthogonality of
the wave functions $\rho_i(w_i)$ leads to integrals over the square
complex norm of each $(\rho_\s,\rho_\d)$-contribution.
But this contains a square delta, so that the norm diverges.
Therefore, the point spectrum
of the operator $\h U_6(v)$ does not contain the zero,
as we wanted to show. Actually, one can apply the same line of reasoning
for any other possible eigenvalue of $\h U_6$, not just for zero,
showing that in fact the point spectrum of this operator is empty.

\subsection{Physical Hilbert space}
\label{s3.3}

Now that we have seen that the solutions to the Hamiltonian constraint are
completely determined by the data on the initial slice $v=\ep$, we can
identify these solutions with the corresponding initial data and
characterize the physical Hilbert space by providing
a Hilbert structure to the data, belonging in
principle to the dual of the vector space
$\text{Cyl}_{\lambda_\s^\star}\otimes
\text{Cyl}_{\lambda_\d^\star} \subset
\mathcal H_{\lambda_\sigma^\star}\otimes\mathcal
H_{\lambda_\delta^\star}$.

In order to endow them with an inner product, we take a(n over) complete set
of classical observables forming a closed algebra, and we impose that the
quantum counterpart of their complex conjugation relations become adjointness
relations between operators. Such a set
is formed by the operators $\widehat{e^{ix_i}}$ and
$\hat{U}_i^{\omega_i}$, with $\omega_i\in\mathcal
Z_{\ep}$ and $i=\sigma,\delta$. For
$\psi(\la_\s,\la_\d)\in \text{Cyl}_{\lambda_\s^\star}\otimes
\text{Cyl}_{\lambda_\d^\star}$ (and for the initial data by duality),
these operators are defined as
\begin{align}
 \widehat{e^{ix_\sigma}}\psi(x_\sigma,x_\delta)&={e^{ix_\sigma}}
 \psi(x_\sigma,x_\delta),\\
\hat{U}_\sigma^{\omega_\sigma}\psi(x_\sigma,x_\delta)&=
\psi(x_\sigma+\omega_\sigma,x_\delta),
\end{align}
and similarly for $\widehat{e^{ix_\delta}}$ and
$\hat{U}_\delta^{\omega_\delta}$. Clearly, all these operators are
unitary in $\mathcal H_{\lambda_\sigma^\star}\otimes\mathcal
H_{\lambda_\delta^\star}$,
according with their reality conditions. Therefore, we conclude that this
Hilbert space is precisely the physical Hilbert space of the vacuum
Bianchi I model.

\section{Hybrid quantization of the Gowdy $T^3$ cosmologies}
\label{s4}

The Gowdy $T^3$ model can be viewed
as homogeneous Bianchi I backgrounds which allow certain inhomogeneous modes
of the gravitational field to propagate along one direction. This natural
separation in homogeneous and inhomogeneous sectors motivated a hybrid
quantization of the model which combines the loop quantization of the
Bianchi I phase space with a natural Fock
quantization for the inhomogeneities, and which was carried out in
Refs.~\cite{hybrid1,hybrid2}
adopting scheme A for the improved dynamics in the quantization of the
Bianchi I sector. This separation of
degrees of freedom is nonperturbative and independent
of the strength of the inhomogeneities at the classical level. Although ideally
one should perform a LQC quantization for the inhomogeneous degrees of freedom
as well, this hybrid approach is justified if the
most relevant quantum geometry effects (but not necessarily {\sl all}
quantum effects) are those that affect the homogenous background so that
one can establish a kind of perturbative hierarchy in their treatment.
In addition, it is natural to adopt a Fock quantization of the
inhomogeneities in this context based on the expectation that a conventional
Fock description of the inhomogeneities ought to be recovered from LQC in a
regime where quantum geometry phenomena are negligible. In this case
there exists a privileged Fock quantization under
certain requirements on the symmetries of the vacuum and on the existence
of a unitary dynamics with respect to an emergent time \cite{gowdy-fock};
these properties provide a natural Fock quantization for the inhomogeneous
modes.

Here, we will show that the hybrid quantization of the
Gowdy model employing scheme B for the loop quantization of the homogeneous
sector is also viable. As in previous works
with the other scheme \cite{hybrid1,hybrid2}, this is not
a trivial issue owing to the coupling between the homogeneous and
inhomogeneous sectors in the Hamiltonian constraint. Moreover, now the
structure of the homogeneous sector is much more complicated, owing to the intricacy of
the holonomy operators in the new scheme B.
The results obtained in Sec.~\ref{s3} will be essential in order to
see that the hybrid quantization is well defined within scheme B as well.
Our demonstration provides a necessary justification for the steps followed in
Ref.~\cite{mbmm} where the
physical Hilbert space of this new hybrid Gowdy model was obtained.

\subsection{Kinematical structure and Hamiltonian constraint operator}
\label{s4.1}

As in the Bianchi I model, since the spatial topology is that of a
three-torus, we have $\t, \s, \d \in S^1$ with a coordinate length of $2
\pi$.
Following a careful gauge-fixing \cite{cmm,hybrid2}, one finds
that the information about the homogeneous degrees of freedom
(which describe the subfamily of homogeneous space-times in the
Gowdy model) can be encoded in the Bianchi I variables $c_i$ and $p_i$
introduced in Sec.~\ref{s2.1}. On the other hand, the inhomogeneities
corresponding to the content of gravitational waves can be
described by a single metric field (without a zero mode), which in turn
can be described by
creation and annihilation-like variables $\{(a_m, a_m^*),\; m \in \mathbb{Z}
- \{0\}\}$, defined in the same way as the natural variables that one would adopt if the
field behaved as a free massless scalar field. Owing to the partial
gauge-fixing, only two global constraints remain on the
system: the zero mode of the Hamiltonian
constraint, which generates time reparametrizations,
and the zero mode of the $\t$-diffeomorphism constraint, which
generates translations around the $\t$-circle (see Ref.~\cite{hybrid2} for
details).

In order to proceed with the hybrid quantization of the Gowdy model, we follow the
LQC approach for the homogeneous degrees of freedom and, as in the Bianchi I
model, we adopt the prescription $p_i c_i \to \h \th_i$
where the operators $\h \th_i$ are defined in Eq.~(\ref{defTh}) while for the
inhomogeneities we promote the creation and annihilation
variables to operators in the standard quantum field theory fashion.
The kinematical Hilbert space is then the tensor
product of the polymer space of the Bianchi I model times the resulting Fock
space for the inhomogeneities \cite{mbmm}.

The generator of translations around the $\t$-circle only affects the
inhomogeneities and it is straightforward to impose in the quantum theory
\cite{hybrid1,hybrid2,mbmm}.
On the other hand, as we pointed out earlier, the Hamiltonian constraint
couples the homogeneous and inhomogeneous sectors in a nontrivial way. The
resulting operator has the explicit form \cite{mbmm}

\begin{align} \label{hamiltonian} \h \mC_H =& - \f{1}{16 \pi G \g^2}
\Big[\h\th_\t \h\th_\s + \h\th_\s
\h\th_\t + \h\th_\t \h\th_\d + \h\th_\d \h\th_\t + \h\th_\s \h\th_\d +
\h\th_\d \h\th_\s\Big] \nonumber \\ & \quad + \f{1}{16 \pi} \left(
\widehat{\left[\frac{1}{|p_\t|^{1/4}}\right]}^2
\f{(\h\th_\s + \h\th_\d)^2}
{\g^2} \widehat{\left[\frac{1}{|p_\t|^{1/4}}\right]}^2 \widehat{\Hi}
+ 32 \pi^2 \widehat{|p_\t|} \widehat{\Ho} \right), \end{align}
where
\be \label{HoHi}
\widehat{\Ho} = \sum_{m \ne 0} |m|  \h{a}_m^\dag \h{a}_m \qquad \mathrm{and}
\qquad
\widehat{\Hi} = \sum_{m \ne 0} \f{1}{2 |m|} \Big(2  \h{a}_m^\dag \h{a}_m
+ \h{a}_m \h{a}_{-m} + \h{a}_m^\dag \h{a}_{-m}^\dag \Big) \ee
are the terms that involve the inhomogeneities and
the regulated ``inverse triad'' operator representing
$|p_\t|^{-1/4}$ is given
by \cite{awe3,mbmm}
\be \widehat{\left[\frac{1}{|p_\t|^{1/4}}\right]}
|v,\la_\s, \la_\d \rangle = \f{\sqrt{2 |\la_\s
\la_\d|}}{(4 \pi \g \sD \lp^3)^{1/6}}\Big(\sqrt{|v+1|} - \sqrt{|v-1|}\Big)
|v,\la_\s, \la_\d \rangle. \ee
Note that the first term in Eq.~\eqref{HoHi} is the Hamiltonian of a
free massless scalar field and the second is a quadratic interaction
Hamiltonian. Owing to the coupling between these terms and those of the
homogeneous sector, \emph{it is not guaranteed that the hybrid approach is
physically feasible beyond the kinematical level,} namely, once the constraints
are imposed. In the remainder of this section we confirm that one attains in fact a
well-defined physical theory.

The explicit form of the action of the Hamiltonian constraint operator on
kinematical states is easy to compute
as most of the terms in the operator have
already been considered in the Bianchi I model or are extensions of
well-known operators.
The most significant subtleties concern the action of
$\widehat{\Hi}$ as this operator includes the sum of all
$\h{a}_m^\dag \h{a}_{-m}^\dag$ terms ($m \neq 0$), each of which creates
an extra pair of ``particles'' in the modes $m$ and $-m$. However,
in spite of the fact that $\widehat{\Hi}$ creates an infinite number of
particles, one can prove that it is a well-defined operator in a
suitable dense domain of the Fock space \cite{mbmm}.
On the other hand, it is straightforward to see that $\h
\mC_H$ leaves invariant Hilbert subspaces which are the tensor product of the
superselection sectors of the Bianchi I model times the Fock space. Therefore,
as in the Bianchi I model, we can restrict the study to separable Hilbert
subspaces whose states have, in the homogeneous sector, quantum numbers
$(v,\la_\s,\la_\d)$
with support in discrete sets contained in the positive octant. Let us remember
that while $v$ takes values in a semi-lattice of constant step equal to $4$ with
a minimum equal to $\epsilon\in(0,4]$, the values of $(\la_\s,\la_\d)$
densely cover the positive quadrant of the real plane.

Even though, from a physical perspective,
one is only interested in small inhomogeneities which produce
a perturbation around the homogeneous Bianchi I background, the hybrid
quantum model is well defined and consistent without restrictions on
the wave numbers or occupation numbers of the modes, and the evolution
can be obtained in much the same manner as in the vacuum Bianchi I model.
Using the result that the Bianchi I model in scheme B leads to a
well posed initial value problem on the section of constant volume
$v=\ep$ from which
one can evolve the physical state in steps of four units in $v$, one can
show that the physical evolution in the hybrid Gowdy model is also (formally)
solvable adopting a perturbative approach, in which the effect of the
interaction term $\widehat{\Hi}$ is treated as small compared to the free
Hamiltonian term $\widehat{\Ho}$. Starting with initial data at $v=\ep$,
one can then find the form of the physical wave functions at $v=\ep+4$
in a perturbative expansion. With this data, one can continue the evolution
to the next section $v=\ep+8$. This procedure can be repeated until one
obtains the expression of the physical wave function at the
wanted value of $v$ and up to the desired perturbative order. Actually,
this perturbative expansion can be understood as an asymptotic expansion in
the limit in which the Immirzi parameter tends to infinity. The details of this
perturbative expansion are presented in Ref.~\cite{mbmm}.

The important point here
is that the evolution is well defined in this perturbative approach. This is
mainly due to the
fact that the initial value problem for the vacuum Bianchi I sector is well posed and
therefore this result depends upon the proof presented in Sec.~III.
In this sense, the
initial value problems in the vacuum Bianchi I model and the Gowdy $T^3$
model are closely related.

\section{Discussion}
\label{s5}

In this paper we have first considered vacuum Bianchi I universes
with a three-torus topology in the framework of LQC, adopting a new scheme
for the improved dynamics which was put forward in Ref.~\cite{awe2}.
We have examined some of the aspects of this quantization which had remained
unanswered in Ref.~\cite{awe2}, like the decoupling of triad components with different
orientations under the action of the Hamiltonian constraint, the structure of
the superselection sectors in the anisotropies, and the evolution of physical
states in terms of the volume as a discrete, internal evolution variable. Then,
we have shown that the initial value problem is well posed and
completed the
quantization of the vacuum Bianchi I
model, following scheme B for the implementation of the improved dynamics.

In Sec.~\ref{s4}, we have used the results regarding the vacuum Bianchi I
model in order to show that the scheme B hybrid quantization of the linearly
polarized Gowdy $T^3$ cosmological model is viable.
This hybrid quantization provides a first step towards a
better understanding of the effect of inhomogeneities in LQC;
this is necessary if one wants to eventually obtain
predictions about the influence and possible traces of quantum
gravity in phenomena like primordial gravitational waves, the cosmic
microwave background, and the physics of the
early universe in general.

The loop quantization of the Bianchi I model leads to superselection
in separable sectors not only for the volume, but also for the
anisotropies. Moreover, every superselection sector
is restricted to an octant. This is because
the Hamiltonian constraint operator, due to appropriate factor ordering choices,
does not mix eigenstates of the densitized triad components with
different orientations. Moreover, while the superselection sectors
in the volume of the Bianchi I universes consist of equidistant points
forming a semi-lattice, the superselection sectors in the anisotropies are
dense sets in the real semi-axis.
On the other hand, the restriction to
a definite orientation of the triad components without imposing
any kind of boundary conditions,
together with the fact that the initial value problem for the evolution
is well posed at the minimum value of $v$ (i.e., $v=\ep$),
can be regarded as a realization of a no-boundary prescription for the
dynamics. In addition, it is worth emphasizing the result
that the discrete evolution in $v$ is well defined starting from
the initial section $v=\ep$. If this were not the case, the evolution would
break down for Bianchi I cosmologies in vacuo for scheme B
and, without any reasonable justification, the inclusion of matter
would turn out to be critical in order for the
dynamics of the model to be viable; we have shown that this is not the case.

Concerning the hybrid quantization of the Gowdy model, an important point
is that the LQC/Fock split that
we have considered assumes that the quantum behavior of the inhomogeneities
can be well approximated by conventional quantum field theory methods so that
any quantum geometry effects due to the presence of these inhomogeneities can be
neglected as perturbatively small. In a true loop quantization of all the
gravitational degrees of freedom (i.e., presumably in a reduction of LQG
by a suitable incorporation of the symmetries of the Gowdy model), one should
treat both the homogeneous and the inhomogeneous sectors in the same manner,
that is to say, all of the degrees of freedom should
be quantized \`a la loop. From this perspective, the main assumption
in our analysis is that the qualitative results of the hybrid
quantization capture the physics of a full loop quantization so long as
the inhomogeneities are not directly affected in a significant way
by quantum geometry phenomena.  It is worth
noticing that similar assumptions are implicit in the treatment of other
models with matter in LQC inasmuch as matter fields are usually quantized by
standard methods rather than by adopting a unified polymer quantization for
all of the degrees of freedom, gravitational or not.

Notice nonetheless that, even at this level, one can see that the hybrid
quantization approach is sufficient to ensure that the classical
cosmological singularities are resolved as the
singular states corresponding to vanishing Bianchi I scale factors
decouple under the action of the Hamiltonian constraint
operator \cite{mbmm,note1}.
In addition, the quantum dynamics are well posed
as one can use $v$ as an evolution variable and, given the
wave function at $v = \ep$, one can derive the wave function for all other
values of $v$ in the same superselection sector in a perturbative
expansion in the interaction term for the inhomogeneities.

There remain many open questions to be addressed, of course, the most
important being a numerical study of the evolution of the wave function.
This is a very difficult task as even the vacuum Bianchi I model in
LQC has not yet been studied numerically in scheme B.  We propose
to begin with a simpler task and study effective equations
associated to the model; this should yield some insight
into the most relevant quantum geometry corrections to the classical model
\cite{mmpwe}.
The most important point, however, is to understand inhomogeneities
in LQC more deeply and to do this one will have to consider more
general inhomogeneous space-times within the framework of LQC
in order to obtain physical predictions about our early universe
and understand their consequences.

\section*{Acknowledgements}

The authors are grateful to L.J.~Garay, D.~Mart\'in de Blas, P.~Moniz,
J.~Olmedo,
T. Paw{\l}owski, and specially to J.M.~Velhinho for enlightening
conversations and discussions. This work
was supported by the Spanish MICINN Project
FIS2008-06078-C03-03, the Spanish Consolider-Ingenio 2010
Program CPAN (CSD2007-00042), the NSF grant PHY0854743,
the George A. and Margaret M. Downsbrough Endowment, the Eberly
research funds of Penn State, Le Fonds qu\'eb\'ecois de la recherche sur la
nature et les technologies, and the Edward A. and Rosemary A. Mebus
funds. M.~M.-B.~is supported by CSIC
and the European Social Fund under the grant I3P-BPD2006.  E.~W.-E.~thanks
the CSIC for their hospitality during his visit there.


\begin{thebibliography}{99}
\bibitem{note0} There is a discrepancy in signs between this
representation and that of Ref.~\cite{awe2} because, in
the latter, $\gamma$ is considered to change
sign under internal parity transformations, namely
$\gamma=|\gamma|\text{sgn}(v)$, while here, as usual, we treat $\gamma$
as a positive free parameter.
\bibitem{note1} In general, other types of singularities do not develop as
far as the field, its spatial derivatives, and its momentum
remain finite.
\end{thebibliography}
\end{document}